\documentclass{aa}
\usepackage{graphicx,natbib}
\usepackage{longtable,lscape}

\begin{document}
\title{New Galactic Star Clusters Discovered in the VVV Survey\thanks{Based on observations gathered with VIRCAM, VISTA of the ESO as part of observing programs 172.B-2002}}
\author{J. Borissova\inst{1} 
   \and
   C. Bonatto \inst{2}
   \and
   R. Kurtev\inst{1}
   \and
   J. R. A. Clarke\inst{1}
   \and
   F. Pe\~{n}aloza\inst{1}
   \and
   S. E. Sale\inst{1,5}
   \and
   D. Minniti \inst{5,17}
   \and
   J. Alonso-Garc\'{i}a\inst{5}
   \and
   E. Artigau\inst{3}
   \and
   R. Barb\'a\inst{16}
   \and
   E. Bica\inst{2}
   \and
   G. L. Baume\inst{4}
   \and
   M. Catelan\inst{5}
   \and
   A. N. Chen\`{e}\inst{1,6}
   \and
   B. Dias\inst{7}
   \and
   S. L. Folkes\inst{1}
   \and
   D. Froebrich\inst{8}
   \and
   D. Geisler\inst{6}
   \and
   R. de Grijs\inst{9,10}
   \and
   M. M. Hanson \inst{17}
   \and
   M. Hempel\inst{5}
   \and
   V. D. Ivanov\inst{11}
   \and
   M. S. N. Kumar\inst{12}
   \and
   P. Lucas\inst{13}
   \and
   F. Mauro\inst{6}
   \and
   C. Moni Bidin\inst{6}
   \and
   M. Rejkuba\inst{15}
   \and
   R. K. Saito\inst{5}
   \and
   M. Tamura\inst{14}
   \and
   I. Toledo\inst{5}
   }
    
\offprints{J. Borissova}

\institute{Departamento de F\'{i}sica y Astronom\'ia, Facultad de Ciencias, Universidad de Valpara\'iso, Av. Gran Breta\~na 1111, Playa Ancha, Casilla 5030, Valpara\'iso, Chile (\email{jura.borissova@uv.cl; radostin.kurtev@uv.cl; paco.stilla@gmail.com; j.clarke@dfa.uv.cl; s.folkes@dfa.uv.cl; s.sale@dfa.uv.cl})
\and
Universidade Federal do Rio Grande do Sul, Departamento de Astronomia CP 15051, RS, Porto Alegre 91501-970, Brazil (\email{charles@if.ufrgs.br}; bica@if.ufrgs.br )
\and
D\'epartement de Physique and Observatoire du Mont M\'egantic, Universit\'e de Montr\'eal, C.P. 6128, Succ. Centre-Ville, Montr\'eal, QC H3C 3J7, Canada (\email{artigau@ASTRO.UMontreal.CA})
\and
Facultad de Ciencias Astron\'omicas y Geof\'{\i}sicas, Instituto de Astrof\'{\i}sica de La Plata, Paseo del Bosque s/n, La Plata, Argentina (\email{gbaume@gmail.com})
\and
Departamento de Astronom\'ia y Astrof\'isica, Pontificia Universidad Cat\'olica de Chile, Av. Vicu\~na Mackenna 4860, Casilla 306, Santiago 22, Chile (\email{dante@astro.puc.cl; mcatean@astro.puc.cl; itoledoc@gmail.com; mhempel@astro.puc.cl;rsaito@astro.puc.cl;jalonso@astro.puc.cl})
\and
Departamento de Astronom\'ia, Casilla 160, Universidad de Concepci\'on, Chile (\email{achene@udec.cl;dgeisler@astro-udec.cl})
\and
Departamento de Astronomia, Universidade de Sao Paulo, Rua do Matao, 1226 - Cidade Universitaria, 05508-900 - Sao Paulo/SP - Brasil (\email{bdias@astro.iag.usp.br})
\and
Centre for Astrophysics \& Planetary Science, The University of Kent, Canterbury, Kent, UK, CT2 7NH (\email{df@star.kent.ac.uk})
\and
Kavli Institute for Astronomy and Astrophysics, Peking University, Yi He Yuan Lu 5, Hai Dian District, Beijing 100871, China (\email{grijs@pku.edu.cn})
\and
Department of Astronomy and Space Science, Kyung Hee University, Yongin-shi, Kyungki-do 449-701, Republic of Korea
\and
European Southern Observatory, Ave. Alonso de Cordova 3107, Casilla 19, Santiago 19001, Chile (\email{vivanov@eso.org})
\and
Centro de Astrof\'{i}sica da Universidade do Porto, Rua das Estrelas, 4150-762 Porto, Portugal (\email{nanda@astro.up.pt} )
\and
Centre for Astrophysics Research, University of Hertfordshire, Hatfield AL10 9AB, UK (\email{phyqpwl@herts.ac.uk})
\and
National Astronomical Observatory of Japan, Osawa 2-21-1, Mitaka, Tokyo 181-8588, Japan (\email{motohide.tamura@nao.ac.jp})
\and
European Southern Observatory, Karl-Schawarzschild-Strasse 2, D-85748 Garching bei Munchen, Germany (\email{mrejkuba@eso.org})
\and
Departamento de F\'{\i}sica, Universidad de La Serena, Cisternas 1200 Norte, La Serena, Chile; Instituto de Ciencias Astron\'omicas, de la Tierra y del Espacio (ICATE-CONICET), Av. Espa\~na Sur 1512, J5402DSP, San Juan, Argentina (\email{rbarba@dfuls.cl})
\and
Department of Physics, University of Cincinnati, Cincinnati, OH 45221-0011, USA (\email{margaret.hanson@uc.edu})
\and
Vatican Observatory, V00120 Vatican City State, Italy \\ 
}

\date{Received; accepted}

\abstract
{VISTA Variables in the V\'{\i}a L\'actea (VVV) is one of the six ESO Public Surveys operating on the new 4-meter Visible and Infrared Survey Telescope for Astronomy (VISTA). VVV is scanning the Milky Way bulge and an adjacent section of the disk, where star formation activity is high. One of the principal goals of the VVV Survey is to find new star clusters of different ages.}
{In order to trace the early epochs of star cluster formation we concentrated our search in the directions to those of known star formation regions, masers, radio, and infrared sources.}
{The disk area covered by VVV was visually inspected using the pipeline processed and calibrated $K_{\rm S}$-band tile images for stellar overdensities. Subsequently, we examined the composite $JHK_{\rm S}$ and $ZJK_{\rm S}$ color images of each candidate. PSF photometry of $15\times15$ arcmin fields centered on the candidates was then performed on the Cambridge Astronomy Survey Unit reduced images. After statistical field-star decontamination, color-magnitude and color-color diagrams were constructed and analyzed.}
{We report the discovery of 96 new infrared open clusters and stellar groups. Most of the new cluster candidates are faint and compact (with small angular sizes), highly reddened, and younger than $5$\,Myr. For relatively well populated cluster candidates we derived their fundamental parameters such as reddening, distance, and age by fitting the solar-metallicity Padova isochrones to the color-magnitude diagrams.}
{}


\authorrunning{J. \,Borissova et al.}
\titlerunning{New Galactic Star Clusters Discovered in the VVV Survey.}

\maketitle
%
%________________________________________________________________

\section{Introduction}

It is a well established fact that the majority of stars with masses $>$0.50\,M$_{\odot}$ form in clustered environments (e.g. Lada \& Lada 2003, de Wit at al. 2005).  Therefore, in understanding the formation, evolution, dynamics, and destruction of star clusters gain insights into the formation, evolution, and dynamics of galaxies.  Estimates indicate that the Galaxy presently hosts 35000 or more star clusters (Bonatto et al. 2006, Portegies Zwart et al. 2010).  However, only about 2500 open clusters have been identified and constitute a sample affected by several well known selection effects. Less than a half of these clusters have actually been studied, and this subset suffers from further selection biases. Around 1300 clusters, mainly in the infrared have been discovered through automatic or semi-automatic searches of large scale survey data products from DSS, 2MASS, DENIS and GLIMPSE (e.g. Bica et al. 2003, Mercer et al. 2005, Froebrich, Scholz \& Raftery 2007, Glushkova et al. 2010).  Expectations are that the new generation of all sky surveys (UKIDSS, the VISTA-based VHS and VVV, and Gaia) will add many more. Indeed, a new Galactic globular cluster candidate has been detected already by Minniti et al. (2011) on the initial VVV bulge images.

VVV is one of the six ESO Public Surveys selected to operate with the new 4-meter VISTA telescope (Arnaboldi et al. 2007). VVV is scanning the Milky Way (MW) bulge and an adjacent section of the mid-plane, where star formation activity is high. The survey started in 2010, and it was granted 1929\,hours of observing time over a five year period. It covers an area of 520\,deg$^2$, and is expected to produce a catalog of $\sim10^9$ point sources (Minniti et al. 2010, Saito et al. 2010).  One of the main goals of the VVV Survey is a study of star clusters of different ages in order to build a homogeneous, statistically significant sample in the direction of the Galactic center, thus complementing recent catalogs which are complete up to only 1\,kpc from the Sun (version 3.1, 24/nov/2010 of the Dias et al. 2002 catalog; see also Lamers et al. 2005;  Piskunov et al. 2008).  A sample thus obtained will be used to: 1.) carry out a census of the MW open clusters projected towards the central parts of the Galaxy and in the southern disk covered by VVV; 2.) to establish the contamination by star-cluster like statistical fluctuations in the background and holes in the dust; 3.) to estimate relative cluster formation efficiency in the MW; 4.) to estimate the role of disruption effects; 5.) to put some constraints on the Initial Mass Function; 6.)  to compare the Galactic open cluster systems with those of the LMC, SMC, and other extragalactic cluster populations; as well as addressing many other questions. The results presented in this paper are part of a larger program aimed at characterizing the hidden star cluster population in the Galaxy (Borissova et al. 2003, 2005, 2006, 2008, 2009, 2010; Ivanov et al. 2002, 2005, 2010, 2011; Kurtev et al. 2007, 2008, 2009; Hanson et al. 2008, 2010; Bica et al. 2003, Dutra et al. 2003, Longmore et al. 2011). Initially, the project was based on cluster candidates identified from 2MASS (Skrutskie et al. 2006) and the {\it Spitzer Space Telescope} Galactic Legacy Infrared Mid-Plane Survey
 Extraordinaire (GLIMPSE, Benjamin et al. 2003) survey. To date we have confirmed three massive star clusters - DBSB\,179 (Dutra et al. 2003), Mercer 30 and Mercer 23 (Mercer et al. 2005) with masses approaching $10^4$ solar masses. We also confirmed several new globular clusters (FSR1735, Froebrich et al. 2007; Mercer\,3, Kurtev et al. 2008; and Mercer\,5, Longmore et al. 2011). 

After the first year of the VVV survey we have at our disposal $JHK_{\rm S}$ images of almost the whole disk and bulge area covered by the survey, and $ZY$ images for the disk area. In this paper we report the first results of our focused search for new star cluster candidates in the disk area covered by VVV. We concentrated our search towards known star forming regions associated with: methanol maser emission; hot molecular cores (Longmore et al. 2009); galactic bubbles outlined by GLIMPSE (Churchwell et al. 2006, 2007); infrared and radio sources in order to trace the early epochs of star cluster formation.

\section{Observations and Data Reduction}

The VIRCAM (VISTA Infrared CAMera; Dalton et al. 2006) is a 16 detector-array 1.65\,deg$^2$ infrared camera. Each $2048\times2048$ detector is sensitive over $\lambda =$0.8--2.5\,$\mu$m, and it delivers images with an average pixel scale of 0.34\,arcsec\,px$^{-1}$.  A single exposure corresponds to a patchy individual ``paw print'' coverage on the sky. To fill the gaps, and to obtain a contiguous image, six shifted paw-prints are combined into a ``tile'' covering 1.5\,deg by 1.1\,deg, which in the case of VVV, are aligned along Galactic $l$ and $b$ respectively. The total exposure time of a single tile is 80\,sec.  To cover the VVV survey area, the disk field is then divided into 152 tiles. The data reduction was carried out in the typical manner for infrared imaging, and detials of the procedure are described in Irwin et al. (2004).  

\section{Cluster Search}

\begin{figure}
\resizebox{\hsize}{!}{\includegraphics{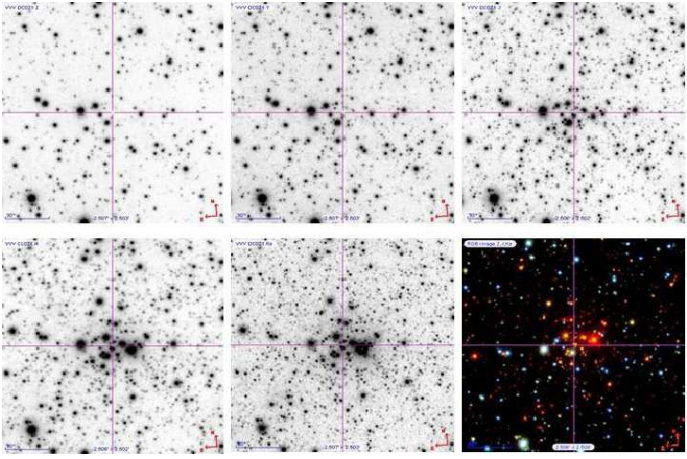}}
\caption{The VVV $ZYJHK_{\rm S}$ images and $ZJK_{\rm S}$ true color image of VVV\,CL036. The field of view is $2.5\times2.5$ arcmin, North is up, East to the left.
}
\label{cl023_all_filters}
\end{figure}

Since we were expecting relatively faint and heavily reddened clusters undiscovered in the 2MASS, DENIS, and the GLIMPSE surveys, we first retrieved the pipeline processed and calibrated $ZYJHK_{\rm S}$ tile images from the Cambridge Astronomical Survey Unit (CASU) VIRCAM pipeline v1.0, Irwin et al. 2004) to visually inspected the $K_{\rm S}$-band tile images for stellar over-densities in the Galactic disk.  We then checked the candidates on the composite $JHK_{\rm S}$ and $ZJK_{\rm S}$ color images. Fig.~\ref{cl023_all_filters} illustrates the process for a typical open cluster candidate, that of VVV\,CL036. As can be seen, the $Z$ and $Y$ images do not contain any over-density of stars, while on the $H$ and $K_{\rm S}$ images the cluster candidate is obvious. The last image in Fig.~\ref{cl023_all_filters} shows the composite $ZJK_{\rm S}$ color image of the cluster and the separation between most probable cluster members (red) and field stars (blue). Prior to the use of color-magnitude diagrams our main criterion to define star cluster candidates were a visually compact appearance, distinctive from the surrounding field and with at least 10 stars with similar colors belonging to this cluster candidate. 

PSF photometry of $15\times15$ arcmin fields surrounding each selected candidate was then performed on the CASU reduced images. We used DAOPHOT-II software (Stetson 1987) within the Image Reduction and Analysis Facility (IRAF\footnote{IRAF is distributed by the National Optical Astronomy Observatories, which are operated by the Association of Universities for Research in Astronomy, Inc., under cooperative agreement with the National Science Foundation}). The PSF was obtained for each frame using approximately a few dozen uncontaminated stars. The typical internal photometric uncertainties vary from 0.005\,mag for stars with $K_{\rm S}$$\sim$13\,mag to 0.15\,mag for $K_{\rm S}$$\sim$18\,mag.  The $J$, $H$ and $K_{\rm S}$ photometry was uniformly calibrated against the 2MASS Point Source Catalog (Skrutskie et al. 2006) generally using several hundred stars in common by least-squares linear regression. For the $Z$ and $Y$ filters we used zero points given in CASU catalogs.  Where possible the saturated stars (usually $K_S \leq 13.5$ mag) were replaced by 2MASS point source catalogue stars.

\section{Field-star decontamination}

In general, poorly-populated clusters, and/or those containing high fractions of faint stars, require field-star decontamination for the identification and characterization of the cluster members. The images of the candidates (Figs.~\ref{cl023_all_filters} and \ref{true_color_1}) clearly show that the decontamination is essential to minimize confusion with the red dwarfs of the Galactic field.

\begin{figure}
\resizebox{\hsize}{!}{\includegraphics{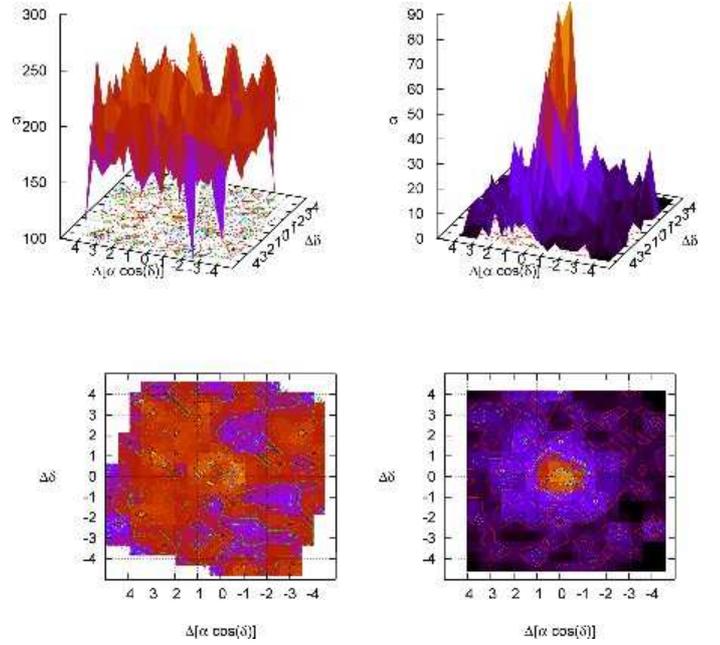}}
\caption[]{Stellar surface-density $\sigma (\rm stars\,arcmin^{-2}$) of VVV \,CL036. The radial density profile produced with the
raw photometry is shown to the left and the right images show the field after statistical decontamination. 
}
\label{cl023_3d}
\end{figure}

\begin{figure}
\begin{center}
\resizebox{9cm}{!}{\includegraphics{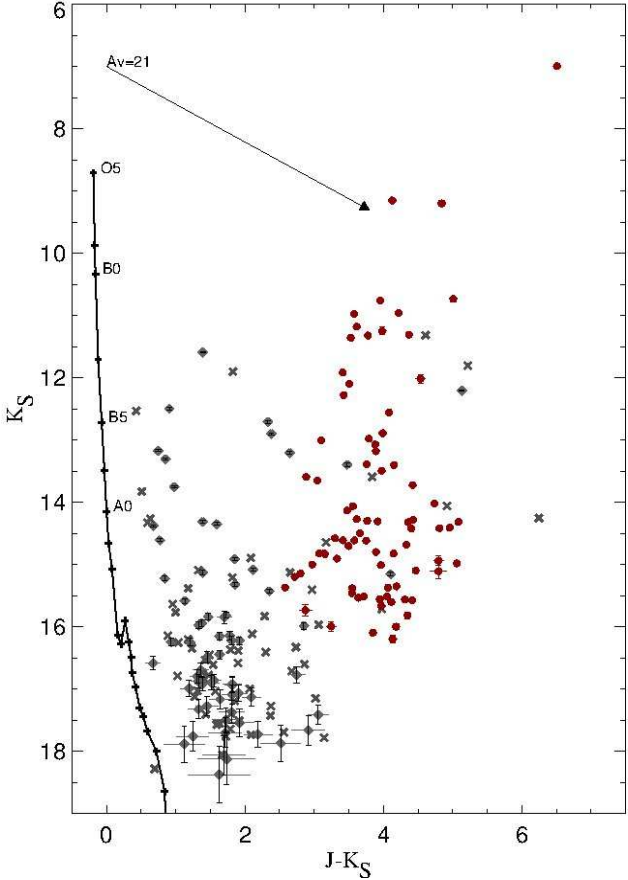}}
\caption[]{VVV observed $(J-K_{\rm S})$ vs. $K_{\rm S}$  CMD extracted from the $R=0\farcm5$ region of VVV CL036. The most probable cluster members are  plotted with filled circles and the filled rombs and crosses stand for the equal area two randomly selected comparison field stars. The continuous  line represents the sequence of the zero-reddening stars of luminosity
class V (Schmidt-Kaler 1982). Reddening vector for $A_{V}$=21\,mag is also shown.}
\label{cl023_cmd}
\end{center}
\end{figure}

For this purpose we used the field-star decontamination algorithm described in Bonatto \& Bica (2010, and references therein), adapted to exploit the VVV photometric depth in $H$ and $K_{\rm S}$. The first step was to define a comparison field that, depending on the distribution of stars, clusters or extinction clouds in an image, may be the form of a ring around the cluster or other different geometry. The algorithm divides the full range of magnitude and colors of a CMD into a 3D grid of cells with axes along $K_{\rm S}$, $(H-$$K_{\rm S})$ and $(J-$$K_{\rm S})$. Initially, cell dimensions were $\Delta$$K_{\rm S}$$=$$1.0$ and $\Delta$$(H-$$K_{\rm S})$$=$$\Delta$$(J-$$K_{\rm S})$$=$$0.2$\,mag, but sizes half and twice those values were also used. We also applied shifts in the grid positioning by $\pm1/3$ of the respective cell size along the 3 axes. Thus, the number of independent decontamination outputs amounted to 729 for each cluster candidate. For each cell, the algorithm estimated the expected number-density of member stars by subtracting the respective field-star number-density\footnote{Photometric uncertainties were taken into account by computing the probability of a star of given magnitude and colors to be found in a any cell (i.e., the difference of the error function computed at the cell's borders).}. Thus, each grid setup produced a total number of member stars $N_{\rm mem}$ and, repeating the above procedure for the 729 different setups, we obtained the average number of member stars $\left<N_{\rm mem}\right>$. Each star was ranked according to the number of times it survived after all runs (survival frequency) and only the $\left<N_{\rm mem}\right>$ highest ranked stars were taken as cluster members. For the present cases we obtained survival frequencies higher than 90\%. Further details about the algorithm are described in Bonatto \& Bica (2010). In Fig.~\ref{cl023_3d}  are shown the large-scale spatial distribution of the stellar surface-densities ($\sigma$, in units of $\rm stars\,arcmin^{-2}$) of VVV\,CL036, built with the raw (left panels), and field-star decontaminated (right) photometry. The decontaminated CMD of VVV\,CL036 is shown in Fig.~\ref{cl023_cmd}.

\section{Catalog of the Cluster Candidates} 
 
Preliminary analysis of the color-magnitude and color-color diagrams reveals 96 previously unknown star cluster candidates or young stellar groups. In Table~\ref{candidates} we list their basic properties. The first column of the table gives the identification followed by: the equatorial coordinates of the cluster candidate's center determined by eye, the number of most probable cluster members after statistical decontamination, eye-ball measured apparent cluster radius in arcsec, the VVV tile name, and comments about the nature of the object. The comments include details taken from the SIMBAD database such as: presence or absence of nebulosity (H\,{\sc{II}} region) around the cluster; known nearby infrared, radio, and X-ray sources; young stellar objects (YSO); outflow candidates and masers.  It should be noted that IRAS positions could have discrepancies of an arcminute or more (particularly if the sources are bright at 60--100\,$\mu$m and not at 12--25\,$\mu$m). Therefore we have listed IRAS sources within 100\,arcsec of our detected clusters (and provided the separations of these with a proximity$\geq$10\,arcsec).
 
\begin{figure}
\resizebox{\hsize}{!}{\includegraphics{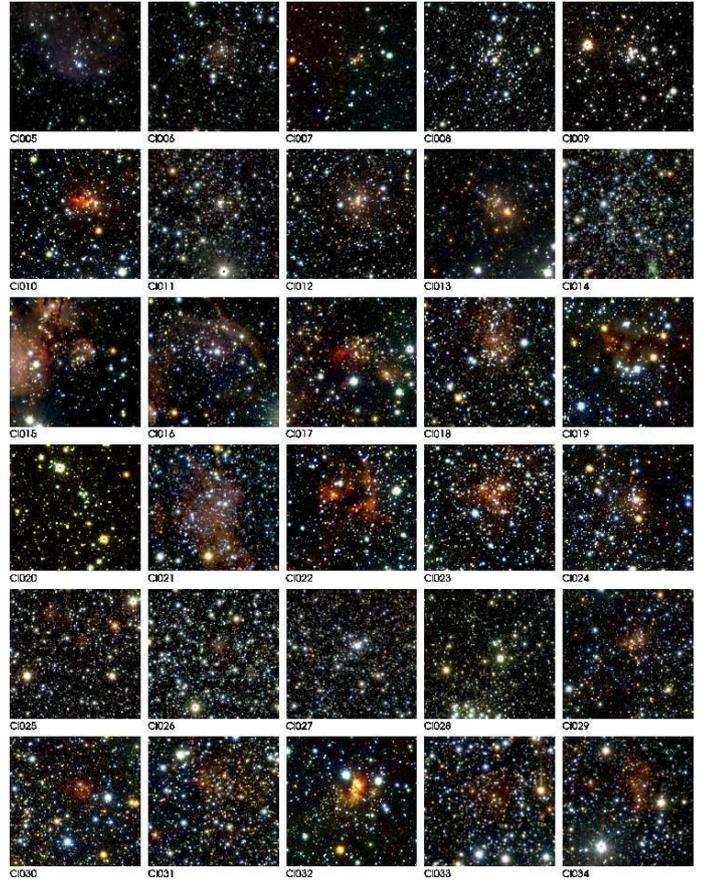}}
\caption{VVV $JHK_{\rm S}$ true color images of VVV open cluster candidates. The field of view is approx. 2.2$\times$2.2 arcmin and North is to the left, East is up.}
\label{true_color_1}
\end{figure}

Fig.~\ref{true_color_1} shows $JHK_{\rm S}$ true color images of some of the newly discovered cluster candidates and stellar groups. The remainder are given in Appendix A.

\begin{figure*}
\resizebox{19cm}{!}{\includegraphics{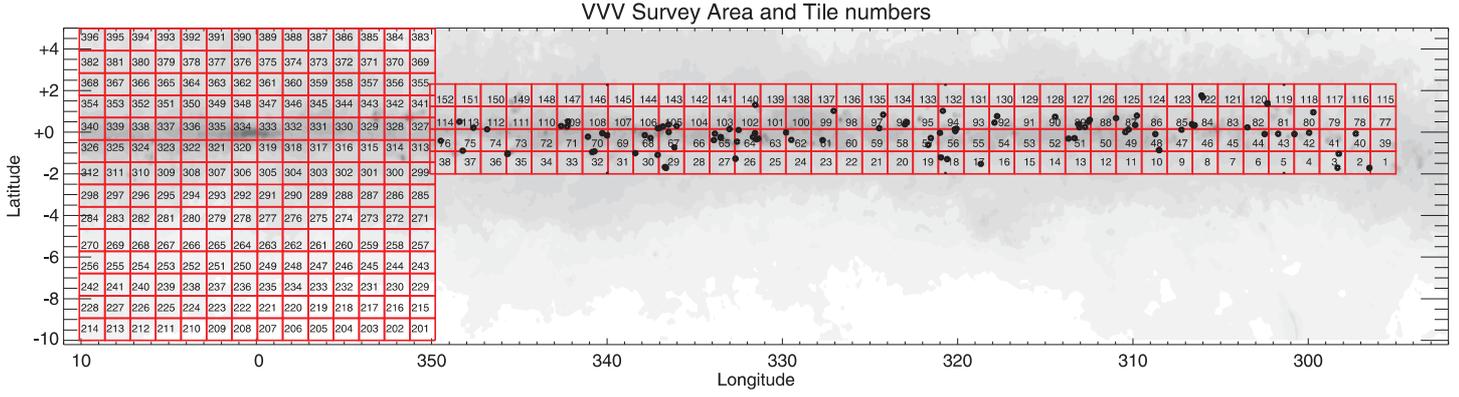}}
\caption[]{ The VVV Survey area, with individual tiles (see Minniti et al. 2010 for details) numbered. This is a plot of Galactic latitude {\it b} version Galactic longitude {\it l}, overplotted on a differential extinction contour map. The positions of the new star cluster candidates from Table~\ref{candidates} are marked.
}
\label{vvv_area}
\end{figure*}

The VVV survey area with the tile numbers is plotted in Fig.~\ref{vvv_area} with the new star cluster candidates from Table~\ref{candidates} overplotted. By the time of this work all disk tiles were available in the CASU database except those of d037 and d133, which represents $99\%$ of the VVV disk area, or 248\,deg$^2$. The spatial distribution of the detected objects on the sky indicates that they are found mainly within Galactic longitude {\it b} $=\pm1.5$\,deg and shows two peaks in Galactic latitude at {\it l} $=$ 310 and 330--340 deg. 

\begin{figure}
\resizebox{\hsize}{!}{\includegraphics{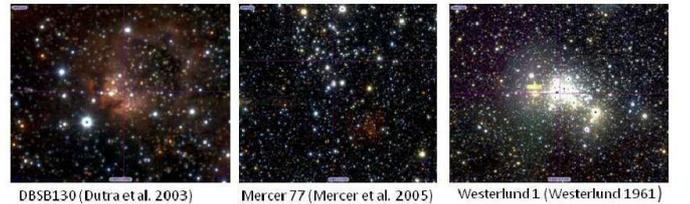}}
\caption{The VVV $JHK_{\rm S}$ true color images of three known massive clusters. The field of view is $2.2\times1.8$ arcmin, 
$4.4\times3.6$ arcmin;  $8.8\times7.2$ arcmin for DBSB\,130, Mercer\,77 and Westerlund 1, respectively.}
\label{known_clusters}
\end{figure}

During the visual inspection we rediscovered 50 of 96 star cluster candidates situated in the VVV disk area from the Dutra at al. (2003) catalog, 65 of 70 star cluster candidates from Mercer et al. (2005), as well as many clusters from WEBDA (Dias et al. 2002) database.  
The VVV true color images of three of them are given in Fig.~\ref{known_clusters}. We did not identify any new cluster candidate similar to 
Westerlund\,1 (right panel).
However, it is hard to estimate the completeness of the catalog presented here with respect to the number of undiscovered clusters in the VVV disk area, as we focused our search towards known radio or mid-IR sources in order to identify young clusters with nebulosity and compact objects.  These can be easily missed by automated search algorithms because the photometry in those regions is affected by abundant extended emission.  Most probably the automated searches will discover many more less-concentrated cluster candidates, although they may still prove unreliable in providing a complete sample.

\begin{figure}
\begin{center}
\resizebox{9cm}{!}{\includegraphics{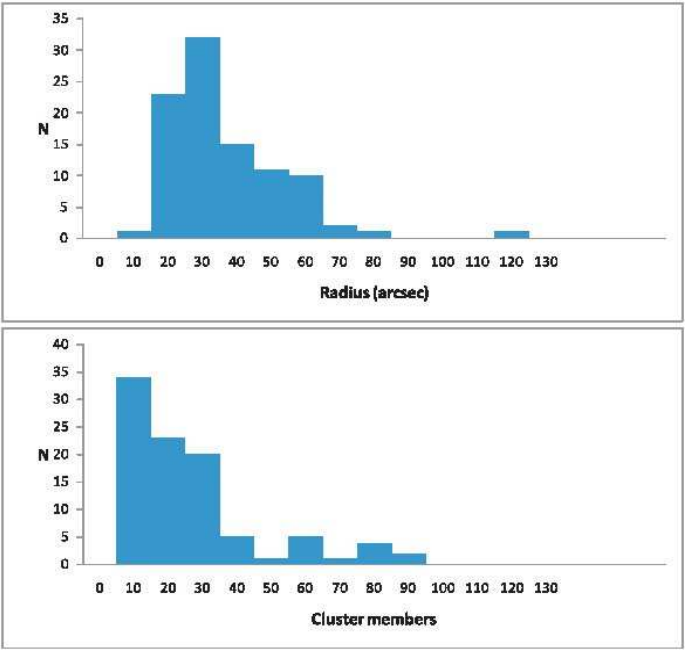}}
\caption{Upper panel: Distribution of the detected objects with measured cluster radius (given in arcsec).
         Lower panel: Histogram of the cluster sample by number of most probable cluster members.}
\label{histograms}
\end{center}
\end{figure}

The cluster radii were measured by eye on the $K_{S}$ 80 second VVV tile images. This method was preferred over automated algorithms, because the majority of our objects are embedded in dust and gas. The area around cluster candidates is smoothed and the density contours are overplotted with the lower limit of the contour equal to the density of comparison field. The mean radius of the sample is $34\pm18$\,arcsec. Here the uncertainty represents the standard deviation of the mean value. This calculated value is smaller than the mean values of $47\pm17$\,arcsec, and $42\pm22$\,arcsec, calculated for the Dutra et al. (2003) and Mercer et al. (2005) clusters respectively for the VVV disk area. The histogram of the number of star clusters vs. 10 arcsec binned radius is shown in Fig.~\ref{histograms}.  It can be seen that most of the clusters have a radius between 20 and 30 arcsec, clearly showing that deep infrared surveys such as VVV allow us to find new faint and compact (with small angular sizes) clusters.

Another possible indication of the richness of the clusters is the number of most probable cluster members, which are given in Column~4 of 
Table~\ref{candidates}. We have to remind the reader however, that the cluster members are selected by statistical decontamination and the 90$\%$ completeness limit of the decontaminated data is $K_{S}$=15.5--16.5\,mag, dependant on crowding and differential reddening. The saturation of the brightest stars additionally complicates the situation, because in some cases it is not possible to replace them directly with 2MASS measurements. Often one 2MASS measurement represents 2 or 3 stars on the VVV images. All above comments have to be considered when determining the distribution of the cluster candidates by the number of most probable cluster members, shown in the lower panel of Fig.~\ref{histograms}. This histogram presents a peak between 10--20 cluster members per cluster. Only 13 clusters have more than 50 members. The small number of cluster members, however, can be also due to the compact nature of the cluster, large distance and/or differential reddening.

\addtocounter{table}{1}

\begin{figure}
\resizebox{\hsize}{!}{\includegraphics{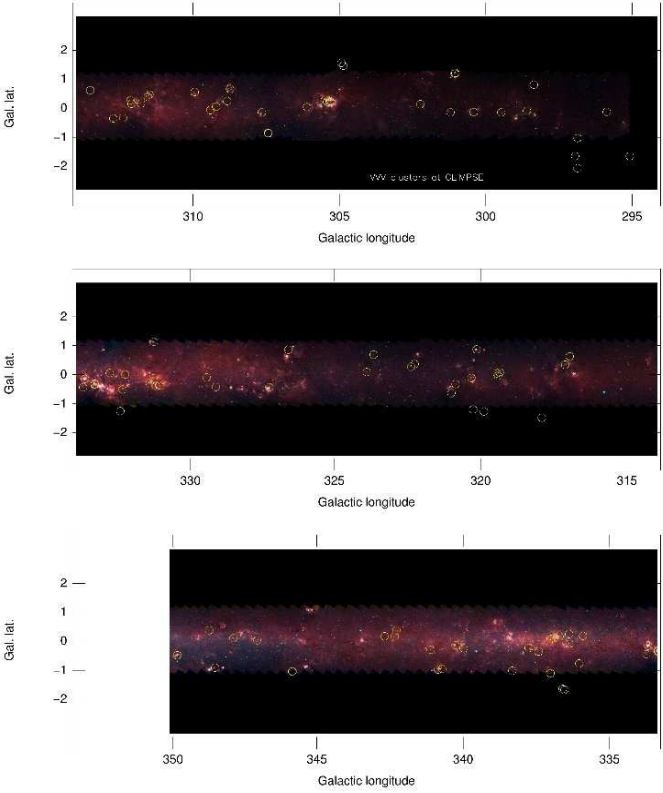}}
\caption[]{ The VVV disk survey area, overplotted on the GLIMPSE 3.6, 4.5 and 8.0\,$\mu$m true color image, with the new star cluster candidates positions.}
\label{vvv_cluster_glimpse}
\end{figure}

The cluster nature of many of the objects listed in Table~\ref{candidates} needs to be confirmed with deeper high-resolution imaging and spectroscopy.  Nevertheless we attempted to make some preliminary classification of the objects based on the morphology of decontaminated color-magnitude and color-color diagrams. Any additional information from the literature was also considered: a nebulosity associated with the object is interpreted as an indicator of youth if it is accompanied by a Main Sequence (MS) and/or Pre-Main Sequence (PMS) on the CMD. The presence of masers, radio, and IR sources within the bounds of the candidate, or of reddened fainter sources that might represent a PMS population, also indicate a young object. Fig.~\ref{vvv_cluster_glimpse} shows a clear correlation between GLIMPSE dust structures and projected position of our star cluster candidates.

According to Plante \& Sauvage (2002) the typical age of embedded clusters is between 1 and 4\,Myr, while the classical open clusters have ages from $10^6$ to $\sim$$10^9$ years. Thus, the candidates with a MS, but lacking any signatures of youth, are considered as classical open clusters with their ages needing to be determined by isochrone fitting. According to the last column of Table~\ref{candidates}, 80 of 96 of our clusters are embedded or very close to nebulosity, while in most of the cases IR, radio, maser, and/or YSO sources are located nearby according to the SIMBAD database. Thus, we infer from the proximity to these sources that $\sim85\%$ of cluster candidates and stellar groups are younger than 5\,Myr. This is in agreement with a Galactic disk active in star formation model.

\section{Parameters of some VVV cluster candidates}

Visible and near-IR data, both imaging and spectroscopy, are normally needed to characterize the main physical parameters of the cluster population (Hillenbrand 1997). Unfortunately, our cluster candidates are practically invisible in the optical. However, follow-up spectroscopy is in progress in the near-IR.  As we have at our disposal the near-infrared color-magnitude and color-color diagrams, we can obtain some initial estimates of basic cluster parameters such as reddening, distance, and age. But we note that this is only particularly effective for the relatively well populated cluster candidates.

\begin{figure}
\begin{center}
\resizebox{10cm}{!}{\includegraphics{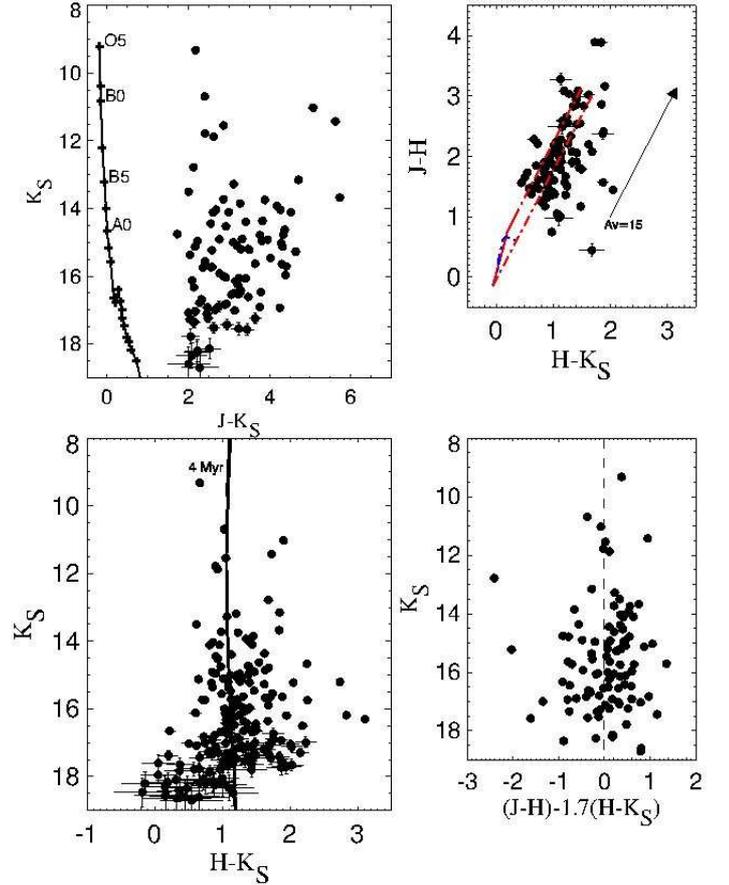}}
\caption[]{VVV observed CMDs extracted from the $R\leq0\farcm8$ region of VVV\,CL062. The top-left panel shows statistically decontaminated most probable cluster members with the Schmidt-Kaler (1982) sequence in the $(J-K_{\rm S})$ vs. $K_{\rm S}$ diagram. The top-right panel gives  $(J-H)$ vs. $(H-K_{\rm S})$ color-color diagram. The continuous lines represents the sequence of the zero-reddening stars of luminosity class\,{\sc{I}} (Koornneef 1983) and class\,{\sc{V}} (Schmidt-Kaler 1982). The reddening vector for $A_{V}=15$\,mag is overplotted and the dotted lines are parallel to the standard reddening vector.  The bottom-left panel shows the $(H-K_{\rm S})$ vs. $K_{\rm S}$ color-magnitude diagram with a 4\,Myr isochrone from Girardi et al. (2010). In the bottom-right panel the reddening free color of $(J-H)-1.70(H-K_{\rm S})$ is plotted vs. $K_{\rm S}$ magnitude.
}
\label{cl028_cmd}
\end{center}
\end{figure}

The mean reddening (interstellar absorption) is well determined from the $J-H$ vs. $H-K_{\rm S}$ color-color diagram. To illustrate the process we show in Fig.~\ref{cl028_cmd} the decontaminated  $J-H$ vs. $H-K_{\rm S}$ color-color diagram of VVV\,CL062. 
%and follow the procedure by Robberto et al.(2010). 
The intrinsic colors of the MS stars (Schmidt-Kaler 1982) and giant branch (Koornneef 1983) are overplotted.  The reddening vectors (Bessell et al. 1998) which encompass the MS stars correspond to a visual extinction of 15 magnitudes.  Clearly, Fig.~\ref{cl028_cmd} indicates that most of the stars are reddened main-sequence stars. The color spread of cluster members, however, is much larger than the typical photometric errors of 0.05 mag, suggesting large differential extinction. The locus lies between the two parallel dotted lines and the calculated mean value and the standard deviation of the fit for this cluster are $A_{V}$=14.7$\pm$0.9 mag. Sources located to the right and below the reddening line may have excess emission in the near infrared (IR-excess sources) and/or may be pre-main sequence stars.
%defined  by the equations $(J-H)-1.83(H-K_{\rm S})+0.098=0$ and $(J-H)-1.83(H-K_{\rm S})+0.50=0$.

Most of the objects in our catalog are young and highly obscured/extinguished clusters (see Table~\ref{candidates}). 
The clusters younger than $\sim30$\,Myr are expected to be affected by differential internal reddening. Indeed, Yadav \& Sagar (2001) show that differential reddening tends to increase towards younger ages, in some cases reaching $\Delta A_{V}$ up to 3\,mag. Unfortunately, we do not yet have enough information to be able to simply estimate the level of differential reddening, as the characteristics of the ISM near the clusters, as well as the profile of extinction between us and the clusters are unknown. If all the extinction is due to a non-uniform thin screen immediately in front of a cluster, then the standard deviation of the extinction to each star (i.e. the measurement of differential reddening) could be as much as twice the mean extinction of stars within the cluster (Fischera \& Dopita 2004). If a cluster is embedded in a cloud which produces much of the extinction along the line of sight to the cluster, we would not only get a contribution to differential extinction from the angular position of each star, but also from its distance, as more distant stars would be, on average, more heavily extinguished.

We attempt to negate the effect of differential reddening by employing the reddening free parameter $Q=(J-H)-1.70(H-K_{S})$, as defined by Negueruela et al. (2007) for OB stars (see also Catelan et al. 2011, for a list of several other reddening-free indices, in the ZYJHK system). We chose this parameter in order to avoid the intrinsic degeneracy between reddening and spectral type (and since we expect to find early OB stars in the majority of the clusters in our sample). The bottom-right panel of Fig.~\ref{cl028_cmd} shows this reddening free parameter vs. $K_{S}$ magnitude. According to Negueruela et al. (2007) the OB stars will have $Q\simeq0.0$ in the diagram, whilst stars with $Q<-0.05$ and large values of $(J-K_{{\rm S}})$ can be classified as infrared excess objects and therefore PMS candidates. The preliminary separation between reddened OB stars and PMS stars can be very useful for analysis of the color-magnitude diagrams and especially for isochrone fitting. 

In general, without the distance of at least 2-3 cluster members having been estimated by spectroscopic parallax and/or knowledge of differential reddening, measurement of the distance and age using only IR photometry is uncertain. Therefore we give only initial estimates for these parameters that can be used for some studies, but that must be confirmed with follow up observations.

\begin{figure}
\hspace{0.5cm}
\resizebox{\hsize}{!}{\includegraphics{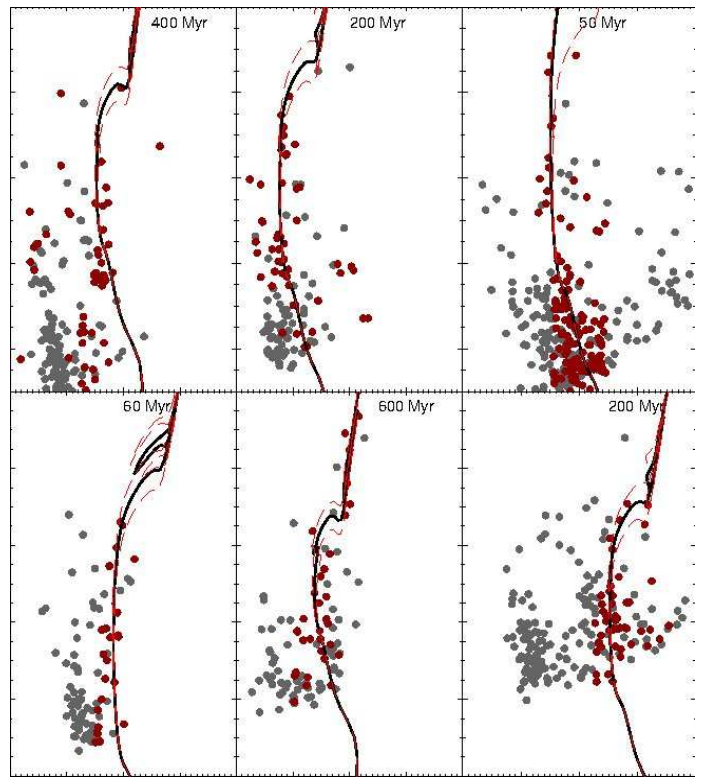}}
\vspace{0.5cm}
\caption[]{VVV CMDs of (from upper left to bottom right) VVV\,CL008, VVV\,CL009, VVV\,CL039, VVV\,CL047, VVV\,CL070, VVV\,CL095. The statistically decontaminated CMDs are shown as red filled circles, the comparison field as gray filled circles. The best isochrone fit (Girardi et al. 2010) is plotted as a solid line, while the dotted lines represent the closest younger and older solutions. 
}
\label{old_cmd}
\end{figure}

The photometric distance and age can be estimated simultaneously by fitting the observed color-magnitude diagram with solar-metallicity Padova isochrones (Girardi et al. 2010) computed with the 2MASS $J$,  $H$, and $K_{s}$ filters. Starting with the isochrones set to zero distance modulus and reddening, we apply shifts in magnitude and color until the fitting statistics reach a minimum value (i.e. difference in magnitude and color of the stars from the isochrone should be minimal). The closest, younger and older fitting solutions were used to provide the age uncertainties. To avoid the dependence of such calculated uncertainties on the resolution of the model grid used, we adopted the largest, most conservative value as the error of age determination. Fig.~\ref{old_cmd} shows the adopted fits superimposed on the decontaminated CMDs for some of the clusters. Parameters derived from the isochrone fit are the true distance modulus $(\rm m-$$\rm M)_{\rm 0}$, age, and reddening $E(J-$$K_{\rm S})$. Reddening estimates can be converted to $E(B-$$V)$ and $A_{V}$ using the equations $E(J-$$K_{\rm S})$$=$$0.56$$\times$$E(B-$$V)$ and $A_{K_{\rm S}}/{A_{V}}=0.118$, which assume $A_{V}$=$3.1$$\times$$E(B-$$V)$ (Dutra et al. 2002). These estimates are presented in Table~\ref{param}.

\begin{table*}
\begin{center}
\caption{Parameters of relatively well populated VVV cluster candidates.}
\label{param}
\begin{tabular}{lrrrr}
\hline\hline
\multicolumn{1}{c}{Name} &
\multicolumn{1}{c}{$A_{V}$} &
\multicolumn{1}{c}{$(m-M)_0$}&
\multicolumn{1}{c}{Dist.} &
\multicolumn{1}{c}{Age}\\
   &\multicolumn{1}{c}{mag}&\multicolumn{1}{c}{mag}& \multicolumn{1}{c}{kpc}& \multicolumn{1}{c}{Myr} \\
\hline
VVV CL008	&	$8.3	\pm0.5$	&	$10.7	\pm0.7$	&	$1.4	\pm0.5$	&	$400\pm50$	\\	
VVV CL009	&	$4.5	\pm0.3$	&	$11.9	\pm0.6$	&	$2.4	\pm0.5$	&	$200\pm40$	\\	
VVV CL039	&	$8.8	\pm0.5$	&	$11.5	\pm0.9$	&	$2.0	\pm0.7$	&	$75\pm40$	\\	
VVV CL041	&	$8.8	\pm0.6$	&	$9.8	\pm0.7$	&	$0.9	\pm0.5$	&	$25-30$	\\	
VVV CL047	&	$10.5	\pm0.9$	&	$14.5	\pm1.3$	&	$7.9	\pm1.3$	&	$60\pm30$	\\	
VVV CL059	&	$20.0	\pm1.4$	&	$10.5	\pm0.8$	&	$1.3	\pm0.5$	&	$20-30$	\\	
VVV CL070	&	$7.2	\pm0.3$	&	$11.5	\pm1.4$	&	$2.0	\pm0.9$	&	$600\pm40$\\	
VVV CL095	&	$13.9	\pm1.9$	&	$12.4	\pm0.8$	&	$3.0	\pm1.4$	&	$200\pm40$	\\	
VVV CL099	&	$13.4	\pm1.6$	&	$9.2	\pm0.7$	&	$0.7	\pm0.6$	&	$20-50$\\	
VVV CL100	&	$14.6	\pm1.2$	&	$13.2	\pm1.4$	&	$4.3	\pm1.1$	&	$5-10$\\	
\hline
\end{tabular}
\end{center}
\end{table*}

\section{Summary} 

In this paper we report the discovery of 96 near-infrared open clusters and stellar groups, found in the Galactic disk area using the ``VVV -- Vista Variables in the V\'{\i}a L\'actea'' ESO Large Survey. Our search concentrated in the directions of known star formation regions, masers, radio, and infrared sources.

The statistics of the foreground reddening of the known Galactic open clusters (e.g. WEBDA, and Fig.~4 of Bonatto et al. 2006) shows that by far the majority of them have $A_{V}$$\leq$3\,mag, with very few having $A_{V}=$5--6\,mag. Table~\ref{param} shows that VVV is really digging deep in the dust, with a mean $A_{V}=11$\,mag and reaching values of $A_{V}=20$\,mag. This highlights the potential of the VVV survey for finding new open clusters, especially those hiding in dusty regions.  Most of the new cluster candidates are faint and compact (with small angular sizes) having radii between 20 and 30 arcsec. An automated search for less-concentrated candidates over the whole VVV area (bulge and disk) based on the color cuts algorithm (Ivanov et al. 2010) is in progress. 

Due to our search being directed towards star formation regions along the Southern Galactic plane we found (based on the preliminary photometric analysis) that approximately 85\% of the star cluster candidates are younger than 5\,Myr. It is hard to esimate the masses of the clusters without accurate distance and age determinations. Taking into account the number of most probable cluster members however, it seems that most of them are intermediate or low mass clusters ($<$$10^{3}$\,$M_{\odot})$.  Spectroscopic follow-up is in progress to verify this, and to derive the spectral types and distances of brighter cluster members.  Once completed, our subsequently constructed homogeneous sample will allow us to better trace the star formation process in the inner Galactic disk.

\begin{acknowledgements}
JB is supported by FONDECYT  No.1080086 and by the Ministry for the Economy, Development, and Tourism's Programa Inicativa Cient\'{i}fica Milenio through grant P07-021-F, awarded to The Milky Way Millennium Nucleus.  The data used in this paper have been obtained with VIRCAM/VISTA at the ESO Paranal Observatory. The VVV Survey is supported by the European Southern Observatory, by BASAL Center for Astrophysics and Associated Technologies PFB-06, by FONDAP Center for Astrophysics 15010003, by the Ministry for the Economy, Development, and Tourism's Programa Inicativa Cient\'{i}fica Milenio through grant P07-021-F, awarded to The Milky Way Millennium Nucleus. This publication makes use of data products from the Two Micron All Sky Survey, which is a joint project of the University of Massachusetts and the Infrared Processing and Analysis Center/California Institute of Technology, funded by the National Aeronautics and Space Administration and the National Science Foundation. This research has made use of the Aladin  and SIMBAD database, operated at CDS, Strasbourg, France. RK acknowledges support from Cento de Astrof\'isica de Valpara\'iso and DIPUV 23/2009. We express our thanks to the anonymous referee for very helpful comments. SLF acknowledges funding support from the ESO-Government of Chile Mixed Committee 2009, and from GEMINI Conicyt grant No. 32090014/2009. DM and DG acknowledge support from FONDAP Center for Astrophysics No. 15010003.  DM is supported by FONDECYT No. 1090213. CB and EB acknowledge support from Brazil's CNPq. JRAC is supported by GEMINI-CONICYT FUND  No.32090002. RdG acknowledges partial research support through grant 11073001 from the National Natural Science Foundation of China. D.G. gratefully acknowledges support from the Chilean Centro de Excelencia en Astrof\'\i sica y Tecnolog\'\i as Afines (CATA).  MSNK is supported by a Ci\^encia 2007 contract, funded by FCT/MCTES (Portugal) and POPH/FSE (EC). JRAC, SES, JAG, FP are supported by the Ministry for the Economy, Development, and Tourism's Programa Inicativa Cient\'{i}fica Milenio through grant P07-021-F, awarded to The Milky Way Millennium Nucleus. ANC received support from Comitee Mixto ESO-GOBIERNO DE CHILE 2009.  ANC and RB are supported by BASAL Center for Astrophysics and Associated Technologies PFB-06. R.S. acknowledges financial support from CONICYT through GEMINI Project No. 32080016. RB acknowledges support from Gemini-CONICYT project 32080001, and DIULS PR09101.

\end{acknowledgements}

\longtabL{1}{
\begin{landscape}%\tabcolsep=1pt
\begin{longtable}{lcccccl}
\caption{VVV Cluster Candidates}\label{candidates}\\
\hline\hline
\multicolumn{1}{l}{Name} &
\multicolumn{1}{l}{RA(J2000)} &
\multicolumn{1}{l}{DEC(J2000)}&
\multicolumn{1}{c}{Numb.} &
\multicolumn{1}{c}{Radius} &
\multicolumn{1}{c}{Tile} &
\multicolumn{1}{c}{Comments }\small\\
&hh:mm:ss&deg:mm:ss& &arcsec  & & \\
\endfirsthead
\caption{continued.}\\
\hline\hline
\multicolumn{1}{l}{Name} &
\multicolumn{1}{l}{RA(J2000)} &
\multicolumn{1}{l}{DEC(J2000)}&
\multicolumn{1}{c}{Numb.} &
\multicolumn{1}{c}{Radius} &
\multicolumn{1}{c}{Tile} &
\multicolumn{1}{c}{Comments }\small\\
&hh:mm:ss&deg:mm:ss& &arcsec  & &\\
\hline
\endhead
\hline
\endfoot   
\hline
VVV CL005	&	11:38:59	&	-63:28:44 	&	25	&	24	&	d001	&	nebulosity, embedded; partofcloud:[SMN83] Lam Cen 1; Be star				\\
VVV CL006	&	11:49:12	&	-62:12:27 	&	12	&	30	&	d039	&	weak nebulosity, embedded; faint; MSX6C G295.7483-00.2076				\\
VVV CL007	&	11:53:51	&	-64:20:30 	&	15	&	20	&	d002	&	weak nebulosity, embedded; IR:IRAS 11513-6403 			\\
VVV CL008	&	11:55:29	&	-63:56:24 	&	25	&	37	&	d002	&	no nebulosity; overdensity				\\
VVV CL009	&	11:56:03	&	-63:18:57 	&	30	&	35	&	d002	&	no nebulosity; overdensity				\\
VVV CL010	&	12:11:47	&	-61:46:23 	&	10	&	28	&	d079	&	strong nebulosity, embedded:GAL 298.26+00.74;IR-[HSL2000] IRS 1;			\\
&	&	&	&	&	& Mas:Caswell CH3OH 298.26+00.74; stellar group; outflow; very red				\\
VVV CL011	&	12:12:41	&	-62:42:31 	&	16	&	6	&	d041 	&	no nebulosity; very concentrated				\\
VVV CL012	&	12:20:14	&	-62:53:04	&	15	&	37	&	d042 	&	nebulosity, embedded; small; IR:IRAS 12175-6236 				\\
VVV CL013	&	12:28:37	&	-62:58:25	&	15	&	27	&	d042 	&	nebulosity, embedded; YSO:G300.3412-00.2190				\\
VVV CL014	&	14:19:09	&	-60:30:46	&	60	&	30	&	d089	&	no nebulosity; overdensity; IR:IRAS14153-6018 (83" away)				\\
VVV CL015	&	12:34:52	&	-61:40:16	&	10	&	20	&	d119	&	nebulosity, embedded; very close to DBSB 77 - part of DBSB 77 or triggered?; \\
&	&	&	&	&	&  IR:Caswell H2O 300.97+01.14	\\
VVV CL016	&	12:35:00	&	-61:41:40 	&	10	&	40	&	d119	&	nebulosity, embedded; close to DBSB 77; young cluster+ SF on the border				\\
VVV CL017	&	12:35:35	&	-63:02:39 	&	10	&	30	&	d043	&	nebulosity, embedded:GAL 301.12-00.20;IR; Mas:Caswell H2O 301.14-00.23				\\
VVV CL018	&	12:44:40	&	-62:47:46	&	25	&	30	&	d082	&	nebulosity, embedded; IR:IRAS 12417-6231 				\\
VVV CL019	&	13:07:06	&	-61:25:03 	&	10	&	50	&	d121	&	nebulosity; YSOs; stellar group 				\\
VVV CL020	&	13:07:36	&	-61:19:28 	&	13	&	24	&	d121	&	no nebulosity; several stars				\\
VVV CL021	&	13:11:51	&	-62:36:52	&	8	&	29	&	d084 	&	weak nebulosity; stellar group; Rad:DWS84 G305.27+0.17				\\
VVV CL022	&	13:12:36	&	-62:37:16 	&	8	&	53	&	d084 	&	nebulosity, embedded; red; Mas:Caswell OH 305.362+00.150				\\
VVV CL023	&	13:13:13	&	-62:33:26	&	15	&	27	&	d084 	&	nebulosity, embedded; IR:IRAS 13100-6217 				\\
VVV CL024	&	13:18:45	&	-62:44:39	&	8	&	27	&	d084 	&	nebulosity, embedded; close to DBSB 85; IR:IRAS 13154-6228A   			\\
VVV CL025	&	13:31:22	&	-63:28:27 	&	5	&	17	&	d047	&	weak nebulosity, embedded; faint; IR:IRAS 13280-6312 (35" away)				\\
VVV CL026	&	13:31:26	&	-63:27:52	&	7	&	18	&	d047	&	weak nebulosity, embedded; faint; IR:IRAS 13280-6312 				\\
VVV CL027	&	13:32:24	&	-62:43:39 	&	40	&	13	&	d047	&	close to nebulosity; overdensity 				\\
VVV CL028	&	13:40:23	&	-61:44:00 	&	10	&	12	&	d086 	&	stellar group; very close to C 1336-614(BH 151); concentrated				\\
VVV CL029	&	13:41:54	&	-62:07:38 	&	20	&	27	&	d086 	&	nebulosity, embedded; IR:IRAS 13384-6152   				\\
VVV CL030	&	13:45:28	&	-62:14:33 	&	 5  &	20	&	d086 	&	nebulosity, embedded;  faint group; IR:IRAS 13419-6159 				\\
VVV CL031	&	13:47:20	&	-62:18:44	&	15	&	45	&	d049	&	weak nebulosity; overdensity; masers; IR:IRAS 13438-6203 				\\
VVV CL032	&	13:50:41	&	-61:35:13	&	15	&	54	&	d087	&	nebulosity:GAL 309.92+00.48, embedded, very red stars; YSO:DZOA 4655-11; DkNeb; 6 Mas				\\
VVV CL033	&	14:03:27	&	-61:16:13 	&	7	&	27	&	d088 	&	nebulosity, embedded; stellar group				\\
VVV CL034	&	14:04:08	&	-61:19:55	&	5	&	34	&	d088 	&	nebulosity, embedded; stellar group				\\
VVV CL035	&	14:06:27	&	-61:29:35 	&	8	&	28	&	d088 	&	nebulosity, embedded; stellar group				\\
VVV CL036	&	14:09:03	&	-61:16:02 	&	52	&	50	&	d088 	&	no nebulosity; overdensity; IR:IRAS 14054-6102				\\
VVV CL037	&	14:09:07	&	-61:24:43	&	15	&	43	&	d088 	&	nebulosity, embedded; IR:IRAS 14054-6110 				\\
VVV CL038	&	14:12:44	&	-61:47:06	&	10	&	20	&	d051	&	nebulosity, embedded; faint; IR:IRAS 14090-6132 				\\
VVV CL039	&	14:15:32	&	-61:41:47	&	72	&	60	&	d051	&	no nebulosity; 2E 1412.0-6127 (X-ray source at 91")				\\
VVV CL040	&	14:44:22	&	-59:11:47	&	20	&	32	&	d092	&	faint; no nebulosity				\\
VVV CL041	&	14:46:26	&	-59:23:17 	&	51	&	54	&	d092	&	no nebulosity; overdensity; IR:IRAS 14428-5911 (92" away)				\\
VVV CL042	&	14:58:48	&	-60:40:07 	&	25	&	74	&	d016	&	no nebulosity; several  bright stars				\\
VVV CL043	&	15:02:56	&	-58:35:55	&	25	&	54	&	d055	&	weak nebulosity; overdensity; close to Mercer 58				\\
VVV CL044	&	15:03:40	&	-58:35:07 	&	10	&	40	&	d093	&	no nebulosity; several bright stars				\\
VVV CL045	&	15:03:47	&	-58:40:11 	&	10	&	54	&	d093	&	no nebulosity; several bright stars; IR:IRAS 14598-5823 (12" away)				\\
VVV CL046	&	15:10:08	&	-58:17:06	&	15	&	20	&	d056	&	nebulosity, embedded; close DBSB 139, triggered?				\\
VVV CL047	&	15:11:52	&	-59:30:30	&	22	&	21	&	d018	&	no nebulosity; faint stars				\\
VVV CL048	&	15:14:01	&	-59:15:13	&	10	&	27	&	d018	&	weak nebulosity, embedded; IR:IRAS 15100-5903				\\
VVV CL049	&	15:14:30	&	-58:11:49	&	68	&	30	&	d056	&	no nebulosity; IR:IRAS 15107-5800 (86" away)				\\
VVV CL050	&	15:21:06	&	-57:57:32	&	12	&	15	&	d057	&	weak nebulosity, embedded; faint stars			\\
VVV CL051	&	15:20:39	&	-56:51:37	&	45	&	50	&	d095	&	no nebulosity; overdensity 				\\
VVV CL052	&	15:21:44	&	-56:52:40	&	21	&	36	&	d095	&	weak nebulosity, embedded; IR:IRAS 15178-5641 				\\
VVV CL053	&	15:27:45	&	-55:48:38	&	38	&	78	&	d096	&	weak nebulosity; several bright stars				\\
VVV CL054	&	15:31:36	&	-56:10:20	&	29	&	54	&	d097	&	weak nebulosity; IR:IRAS 15277-5600 				\\
VVV CL055	&	15:43:36	&	-53:57:52 	&	10	&	15	&	d098	&	weak nebulosity; of?:EGO G326.61+0.80, stellar group				\\
VVV CL056	&	15:52:38	&	-54:34:38	&	10	&	27	&	d061 	&	nebulosity; small; very close to DBSB 146, triggered?				\\
VVV CL057	&	16:02:11	&	-53:22:37	&	10	&	14	&	d062	&	nebulosity; small; IR:IRAS 15583-5314, stellar group				\\
VVV CL058	&	16:02:19	&	-52:55:28	&	20	&	28	&	d062	&	nebulosity, IR:IRAS 15584-5247; Mas; YSOcand				\\
VVV CL059	&	16:05:52	&	-50:47:49	&	35	&	45	&	d102	&	nebulosity; IR:IRAS 16021-5039   				\\
VVV CL060	&	16:11:23	&	-51:42:49	&	10	&	48	&	d064 	&	nebulosity:[KC97c]; Rad:G331.3-00.2, GAL 331.26-00.19; stellar group				\\
VVV CL061	&	16:11:28	&	-52:01:33	&	8	&	22	&	d063 	&	nebulosity; stellar group; IR:IRAS 16076-5154 				\\
VVV CL062	&	16:12:08	&	-51:58:08	&	74	&	39	&	d063 	&	nebulosity; IR:IRAS 16082-5150 				\\
VVV CL063	&	16:12:42	&	-51:45:03	&	10	&	21	&	d064 	&	nebulosity; IR:IRAS 16089-5137 				\\
VVV CL064	&	16:15:18	&	-50:56:48	&	30	&	28	&	d102	&	weak nebulosity; faint				\\
VVV CL065	&	16:17:31	&	-50:32:30	&	25	&	32	&	d103	&	nebulosity:IR:IRAS 16137-5025; YSO:[MHL2007] G332.7673-00.0069 1; very close to Mercer 77 				\\
VVV CL066	&	16:17:59	&	-51:15:10	&	7	&	49	&	d064 	&	nebulosity; small; overdensity; IR:IRAS 16141-5107				\\
VVV CL067	&	15:10:36	&	-57:54:41.77	&	30	&	30	&	d094	&	no nebulosity; concentrated				\\
VVV CL068	&	16:21:28	&	-50:26:24	&	8	&	10	&	d065 	&	nebulosity:GAL 333.29-00.37; stellar group, at the border of DBSB 165  triggered?				\\
VVV CL069	&	16:21:34	&	-50:27:29	&	13	&	60	&	d065 	&	nebulosity; SFR?				\\
VVV CL070	&	16:21:48	&	-51:44:11	&	30	&	26	&	d026	&	no nebulosity; overdensity				\\
VVV CL071	&	16:22:16	&	-50:04:30	&	14	&	25	&	d065 	&	at the border of strong nebulosity, very close to [BDB2003] G333.60-00.21				\\
VVV CL072	&	16:23:49	&	-50:14:20	&	19	&	58	&	d065 	&	nebulosity; IR:IRAS 16200-5007 				\\
VVV CL073	&	16:30:24	&	-48:13:06	&	38	&	20	&	d105	&	no nebulosity				\\
VVV CL074	&	16:32:06	&	-47:49:32	&	87	&	33	&	d105	&	no nebulosity; overdensity; Dark Cloud SDC G336.381+0.190				\\
VVV CL075	&	16:33:30	&	-48:03:35 	&	12	&	27	&	d067	&	weak nebulosity:GRS 336.37 -00.13; IR:MSX5C G336.3618-00.1373; Mas:[HLB98] SEST 107 				\\
VVV CL076	&	16:33:48	&	-47:38:49	&	10	&	20	&	d105	&	5-6 bright stars with the same color				\\
VVV CL077	&	16:34:48	&	-47:32:49	&	21	&	15	&	d105	&	nebulosity; IR:IRAS 16311-4726 				\\
VVV CL078	&	16:35:09	&	-48:46:24	&	22	&	41	&	d067	&	nebulosity; IR:IRAS 16313-4840 -nebulosity				\\
VVV CL079	&	16:35:22	&	-47:28:33	&	18	&	15	&	d105	&	no nebulosity, stellar group, X:SSTGLMC G337.0012+00.0305				\\
VVV CL080	&	16:38:56	&	-47:27:01	&	22	&	25	&	d068	&	weak nebulosity; IR:IRAS 16352-4721 				\\
VVV CL081	&	16:39:43	&	-47:06:57	&	29	&	10	&	d068	&	no nebulosity; very red stars, or dust window?				\\
VVV CL082	&	16:40:39	&	-48:16:07	&	30	&	34	&	d030	&	weak nebulosity, embedded; IR:2MASS J16412047-4900172 			\\
VVV CL083	&	16:41:19	&	-49:00:42 	&	30	&	46	&	d029	&	nebulosity, embedded; YSO:[MHL2007] G336.5299-01.7344 2, \\
&	&	&	&	&	&  [MHL2007] G336.5299-01.7344 3, 2MASS J16412047-4900172  				\\
VVV CL084	&	16:41:24	&	-48:56:33 	&	15	&	50	&	d029	&	no nebulosity, but close; several bright stars				\\
VVV CL085	&	16:45:26	&	-47:13:02	&	37	&	40	&	d068	&	no nebulosity 				\\
VVV CL086	&	16:48:15	&	-45:26:06   &	72	&	35	&	d070 	&	no nebulosity; overdensity; MX5C G340.0160-00.3041			\\
VVV CL087	&	16:48:50	&	-45:09:32	&	10	&	60	&	d070 	&	nebulosity; small overdensity				\\
VVV CL088	&	16:52:34	&	-44:36:07	&	17	&	12	&	d070 	&	close to nebulosity; very concentrated; 5 YSO 				\\
VVV CL089	&	16:53:47	&	-43:16:03	&	83	&	34	&	d109 	&	close to nebulosity;  overdensity				\\
VVV CL090	&	16:54:03	&	-45:18:53	&	8	&	14	&	d070 	&	nebulosity; stellar group; YSO:[MHL2007] G340.7455-01.0021;  close to DBSB 106 				\\
VVV CL091	&	16:54:39	&	-45:14:09	&	10	&	80	&	d070 	&	weak nebulosity; overdensity 				\\
VVV CL092	&	16:54:56	&	-43:21:46	&	10	&	27	&	d109 	&	nebulosity:[WHR97] 16513-4316A ; stellar group				\\
VVV CL093	&	16:56:03	&	-43:04:47	&	9	&	28	&	d109 	&	strong nebulosity; Rad:IRAS 16524-4300, Rad:GBM2006] 16524-4300A; Mas:[SRM89] 16524-4300				\\
VVV CL094	&	17:07:54	&	-40:31:38.6	&	20	&	20	&	d074	&	nebulosity, embedded; Rad:GPSR 346.077-0.055; IRAS 17043-4027				\\
VVV CL095	&	17:10:55	&	-39:41:49 	&	52	&	30	&	d074	&	no nebulosity; overdensity; Dark Cloud SDC G347.082-0.011				\\
VVV CL096	&	17:11:41	&	-41:19:03 	&	10	&	17	&	d036	&	weak nebulosity, embedded; faint				\\
VVV CL097	&	17:11:46	&	-41:18:13 	&	10	&	20	&	d036	&	weak nebulosity, embedded; faint; IR:IRAS 17082-4114 				\\
VVV CL098	&	17:13:06	&	-38:59:45	&	13	&	20	&	d113	&	nebulosity:IRAS 17096-3856; IR:MSX5C G347.9026+00.0486; Rad:GPSR 347.901+0.048; \\
&	&	&	&	&	&  Mas:Caswell CH3OH 347.90+00.05 				\\
VVV CL099	&	17:14:26	&	-38:09:51	&	52	&	30	&	d114	&	no nebulosity; X:CXOU J171424.4-380959 				\\
VVV CL100	&	17:19:15	&	-39:04:34	&	22	&	20	&	d075	&	nebulosity:IRAS 17158-3901; Mas, Rad:GBM2006] 17158-3901; stellar group				\\
%\hline
\end{longtable}	 							
\end{landscape}
}

\appendix
\onecolumn{}  
\section{Three-Color $JHK_{\rm S}$ Composite Images of the Cluster Candidates.}
\begin{figure}[h]
\resizebox{17cm}{!}{\includegraphics{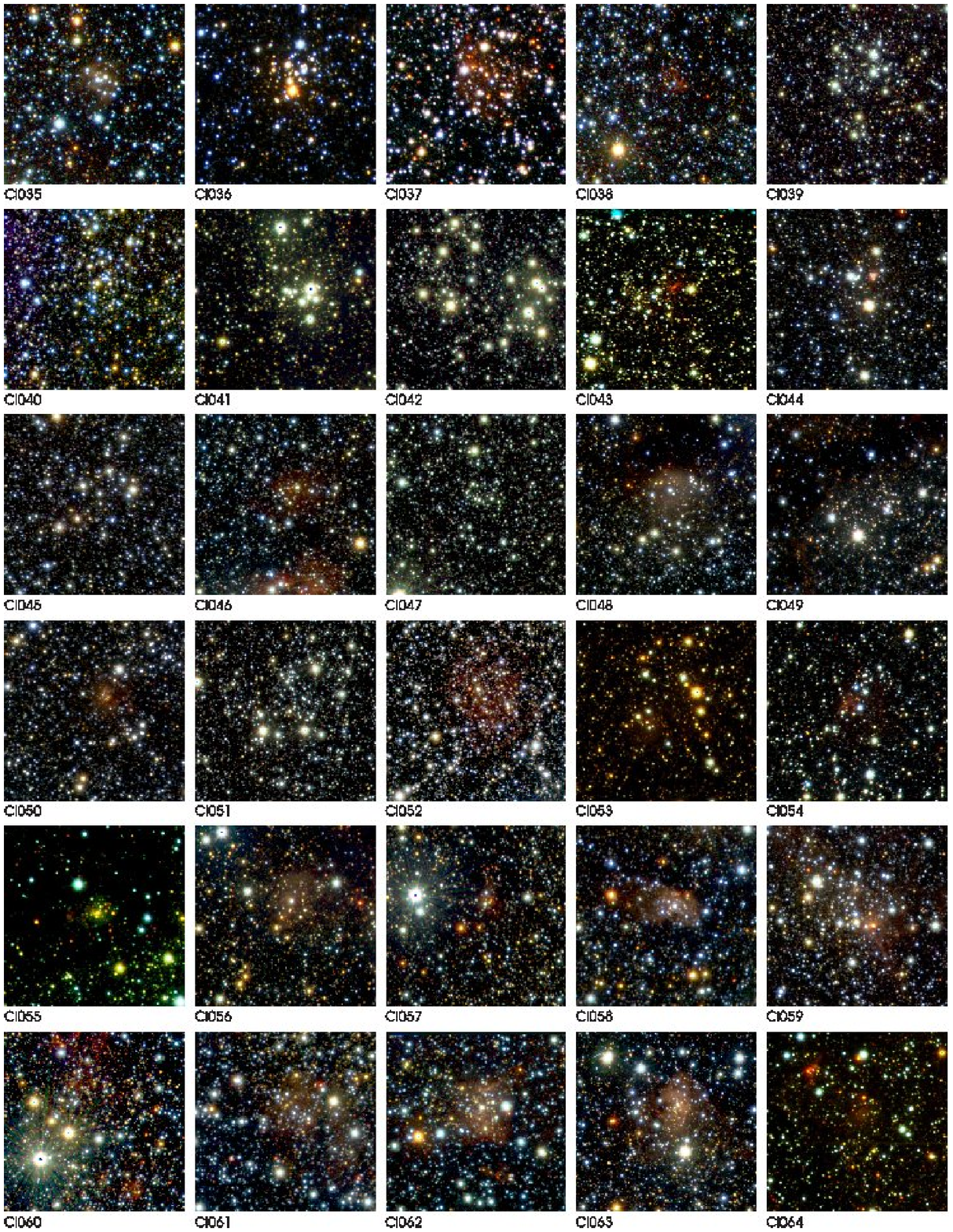}}
%\resizebox{15cm}{!}{\includegraphics{appendix_fig3}}
\end{figure}

\begin{figure*}[h]
\resizebox{\hsize}{!}{\includegraphics{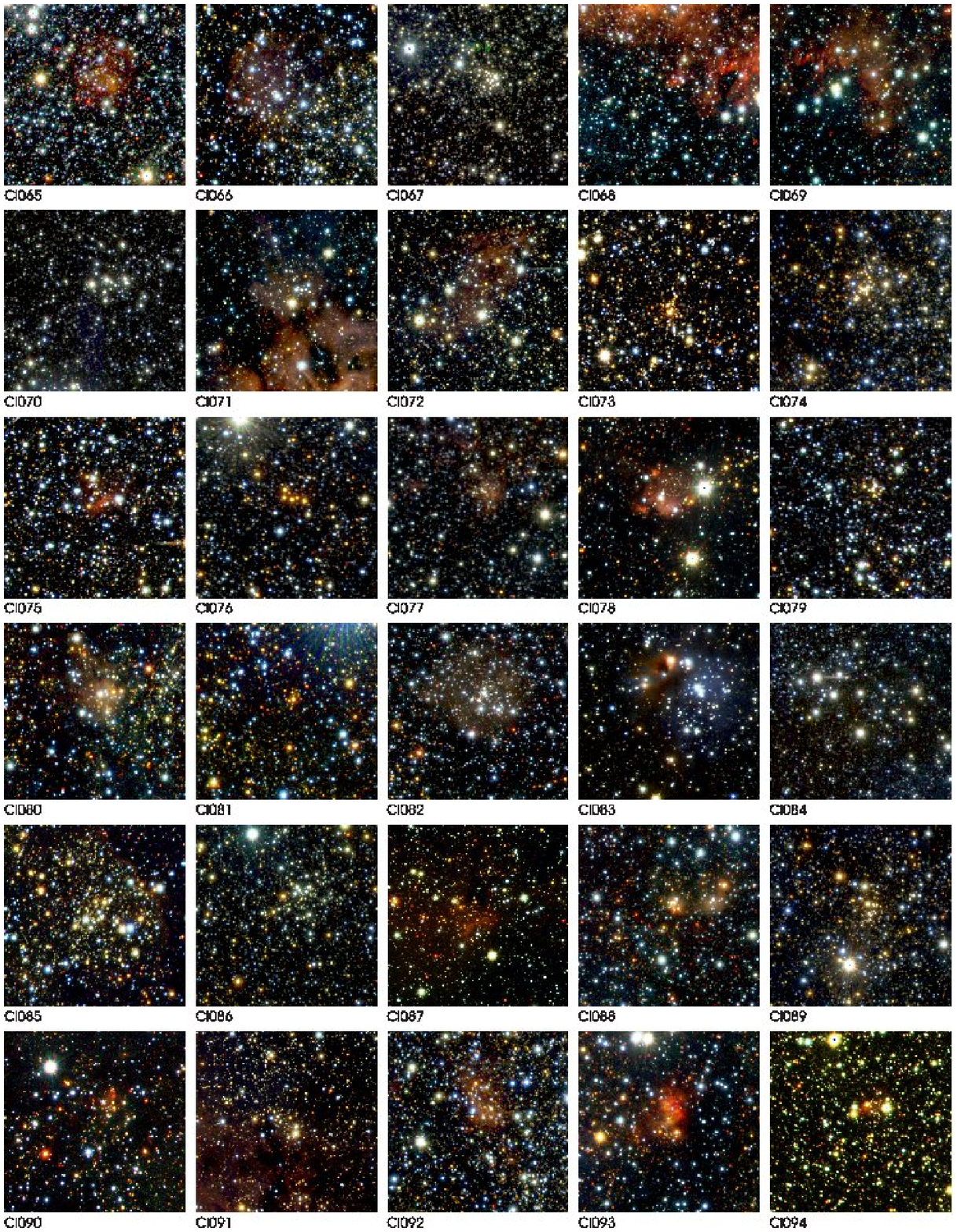}}
%\resizebox{15cm}{!}{\includegraphics{true_color_3}}
\end{figure*}

\begin{figure*}[h]
\resizebox{\hsize}{!}{\includegraphics{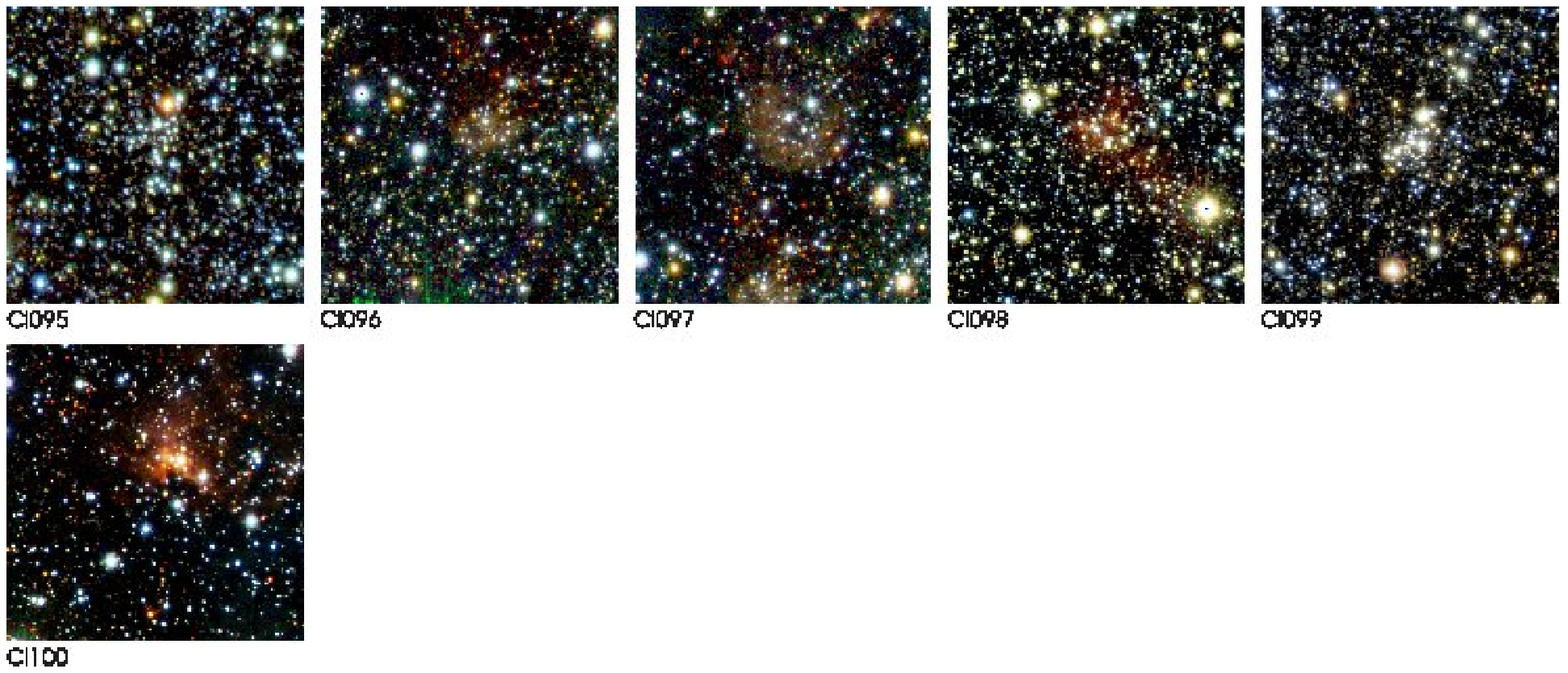}}
%\resizebox{15cm}{!}{\includegraphics{true_color_3}}
\end{figure*}

\end{document}